%Paper: hep-th/9210019
%From: TEMPLE@Vax2.Concordia.CA
%Date: Sun, 4 Oct 1992 15:11 EDT

\magnification\magstep1
\hsize=16.5truecm
\vsize=24truecm
\nopagenumbers
\centerline{\bf On the quantisation of $SU(2)$ magnetic monopole dynamics}
\vskip 0.3in
\centerline{M. Temple-Raston}
\vskip 0.2in
\centerline {Department of Mathematics and Statistics,}
\vskip 0.1in
\centerline{Concordia University, 7141 Sherbrooke Street West,}
\vskip 0.1in
\centerline {Montr\'eal, Qu\'ebec H4B 1R6 Canada}
\vskip 0.5in
\bigskip
\noindent
The low-energy classical dynamics of BPS $SU(2)$ magnetic monopoles can
be modeled using the geodesic approximation of N. Manton [M].  In this
scheme one makes use of the moduli space of static, finite-energy,
non-singular solutions $(A,\Phi)$ to the Bogomol'nyi equations on
${\bf R}^3$,
$${1\over 2}\epsilon_{ijk}F^A_{jk}=D^A_i\Phi.$$
The Lie algebra-valued gauge potential $A$ and Higgs field $\Phi$ are both
in the adjoint representation of $SU(2)$.
$D^A$ is the exterior covariant derivative
of the gauge potential, $A$, and $F^A$ is the gauge curvature.  Finite-energy,
non-singular solutions to the Bogomol'nyi equations are called BPS $SU(2)$
magnetic monopoles, and can be classified by a topological number interpreted
as the magnetic charge.  The geodesic approximation maintains that the
low-energy dynamics of two interacting magnetic monopoles is well-approximated
by the geodesic motion on the moduli space of all BPS magnetic monopoles
with a magnetic charge of two [M].  Fixing the centre-of-mass of the two
magnetic monopole system and factoring out by an irrelevant
overall phase, the moduli space $M^o_2$ is of real dimension four.
Atiyah and Hitchin made use of a natural, non-trivial $SO(3)$ action
on the moduli space $M^o_2$ to determine the unique, complete metric
structure on $M^o_2$ [AH].  The metric can be put into the form
$$ds^2=f(r)^2\,dr^2+a(r)^2\sigma_1^2+b(r)^2\sigma_2^2
+c(r)^2\sigma_3^2.\eqno{(1)}$$
$\{\sigma_1,\sigma_2,\sigma_3\}$ is the usual dual basis for $so^*(3)$.
There is a certain amount of freedom in choosing $f(r)$, we follow
Atiyah and Hitchin and define $f(r)\equiv -abc$.
Then, the equations of motion arising from the analysis in [1] and
which form the basis of our study are:
$$\eqalign {{2bc\over f}{da\over dr}&= (b-c)^2-a^2\ \ \ \ (cyclic),\cr
{dM_1\over dt}&=\bigg({1\over b^2}-{1\over c^2}\bigg)M_2M_3
\ \ \ \ (cyclic),\cr
{d^2r\over dt^2}=-{1\over f}{df\over dr}\bigg({dr\over dt}\bigg)^2+&
{1\over f^2}\bigg({1\over a^3}{da\over dr}M_1^2+
{1\over b^3}{db\over dr}M_2^2+{1\over c^3}{dc\over dr}M_3^2\bigg).}
\eqno{(2)}$$
The first set of equations define the metric coefficients in (1), and
the next three equations are the geodesic equations on the moduli space
$M^o_2$.  The work in [M] and [AH] permit one to examine in some detail
the soliton interaction dynamics contained within the prototype of the
standard model: Yang-Mills-Higgs theory.  Note also that the
geodesic equations above form a non-integrable generalization [TR3]
to the familiar Euler-Poinsot equations for a rigid body---therefore
equations (2) could well be of rather broad interest.

\smallskip
Having now defined in (2) the low-energy classical dynamics of two $SU(2)$
magnetic monopoles, we turn to the quantum dynamics.  We shall assume that
any quantisation of the classical dynamics adheres to the correspondence
principle.  In particular, the appropriate non-relativistic
Schr\"odinger equation for two monopole quantum dynamics must reproduce
the low-energy classical dynamics above in the short wavelength
limit ($h\to 0$).  Gibbons and Manton have shown [GM] that by
taking the Schr\"odinger operator to be proportional to the covariant
Laplacian defined using the metric on the moduli space, $M^o_2$, the
correspondence principle is valid.  Unfortunately it appears that
the quantum wave mechanics arising from this prescription is very
difficult to solve explicitly (see [S] for a partial wave analysis).
While the Schr\"odinger equation proposed by Gibbons and Manton is indeed
very natural, our approach in this short contribution will be to use
the correspondence principle to reduce to the semi-classical domain.
Since, by assumption, all possible quantisation prescriptions of the
$SU(2)$ magnetic monopole must contain the same semi-classical domain,
we have not committed ourselves to any particular Schr\"odinger equation.
The low-energy classical scattering behaviour of $SU(2)$ magnetic monopoles
is sufficient well-understood that we can attempt to compute the
semi-classical differential scattering cross-section.  We do this next.

\smallskip
There are two totally geodesic (real) surfaces, $\Sigma_1$ and $\Sigma_{12}$,
in the moduli space, $M^o_2$, for which the scattering behaviour is
analytically well-understood [AH].  The geodesic motion on $\Sigma_{12}$ is of
more interest to us here, so we shall focus on this surface.  Geometrically,
the surface $\Sigma_{12}$ looks like a funnel asymptotic to a cone of vertex
angle $\pi/3$ at one end, and at the other to a cylinder of radius
$\sqrt{2}/2$.  It is a surface of revolution and therefore $O(2)$-invariant.
Restricting our attention to scattering initial conditions, Atiyah and Hitchin
define a new variable, $\epsilon$, which is related to the impact parameter,
$\lambda$, by $\lambda=1+\epsilon$ [AH], and compute the highest order term
for the scattering angle, $\theta$, as a function of $\epsilon$, given by
$$\theta(\epsilon)\sim\pi\bigg({\epsilon\over 2}\bigg)^{-{3\over 2}},
\eqno{(3)}$$
for $0<\epsilon\le\pi/2-1$.  We may write
$\theta\equiv\theta_o+2\pi j\sim\pi(\epsilon_j/2)^{-3/2}$ where
$0<\theta_o\le\pi$, $j\in{\bf Z}\cup\{0\}$, and solve for
$\epsilon_j$:
$$\epsilon_j\sim 2\bigg({\pi\over \theta_o+2\pi j}\bigg)^{2\over 3}.
\eqno{(4)}$$
This can be used to construct the semi-classical differential
cross-section.  Assuming that $\theta_o$ is away from classical rainbows,
the semi-classical scattering amplitude is [P]
$$f(\theta_o)=\sum_j\sqrt{c_j}e^{i(S_j/\hbar-\pi\mu_j/2)}.
\eqno{(5)}$$
The sum runs over all classical scattering trajectories, $\Gamma_j$,
scattering into $\theta_o$.  From (3) we see that on $\Sigma_{12}$ there
are an infinite number of classical trajectories with scattering angle
$\theta_o$ for BPS $SU(2)$ monopole dynamics.  In fact, we shall now
restrict our sum in (5) to only those scattering trajectories $\Gamma_j$ on
$\Sigma_{12}$.  Of course, depending on $\theta_o$, there may be other
contributing paths not on $\Sigma_{12}$, but excluding them now will
not effect our argument below.  The Maslov index, $\mu_j$, counts the number
of caustics in $\Gamma_j$; $\mu_j=0$ for all trajectories on $\Sigma_{12}$.
The Action $S_j$ in (5) is given by $-\int_{\Gamma_j}{\bf p}.\,d{\bf q}$.
Finally, $c_j$ is the contribution to the scattering amplitude of the
trajectory $\Gamma_j$ with $\epsilon=\epsilon_j$ where $c_j$ is given by
(using (3) and (4))
$$c_j\equiv\bigg\vert{d\theta\over d\epsilon}\bigg\vert^{-1}(\epsilon_j)
\sim{4\over 3\pi}\bigg({\epsilon_j\over 2}\bigg)^{5\over 2}=
{4\over 3\pi}\bigg({\pi\over\theta_o+2\pi j}\bigg)^{5\over 3}.\eqno{(6)}$$
Numerical simulation [TRA] involving all the classical paths of the
Atiyah-Hitchin equations in (2), that is, not just the trajectories on
$\Sigma_{12}$, suggests that there are no classical rainbows round
$\theta_o=\pi/2$, and therefore we can view (5) as being evaluated round
$\pi/2$.  We note that a know rainbow at $\theta=\pi/3$ was successfully
identified by the numerical procedure in [TRA].  From equation (5) the
differential cross-section at $\theta=\theta_o$ is seen to be
$${d\sigma\over d\Omega}=\vert f(\theta_o)\vert^2\approx\sum_j^\infty c_j
+2\sum_{i<j}^\infty\sqrt{c_ic_j}\exp{i\big((S_i-S_j)/\hbar-
\pi(\mu_i-\mu_j)/2\big)}.
\eqno{(7)}$$
The Riemann-Lebesgue Lemma and equation (6) imply that in the classical
limit ($\hbar\to 0$) the differential cross-section is
$${d\sigma\over d\Omega}\approx\sum_j^\infty c_j\sim
\sum_j^\infty{4\over 3\pi}
\bigg({\pi\over\theta_0+2\pi j}\bigg)^{{5\over 3}}.\eqno{(8)}$$
The right hand side of equation (8) is absolutely
convergent, and therefore converges.  Thus the classical differential
cross-section as approximated in (8) is well-defined.  However, it is easy
to see that the second term in (7) is {\it not} absolutely convergent.
For a fixed value of $i$ in (7), the semi-classical scattering amplitude
is only conditionally convergent since the series
$$\sum_j^\infty\vert\sqrt{c_j}\vert\sim\sum_j^\infty{2\over\sqrt{3\pi}}
\bigg({\pi\over\theta_o+2\pi j}\bigg)^{5\over 6},$$
is a divergent hyperharmonic series for all values of $\theta_o$.
Therefore, if the differential cross-section (7) converges, it is only
conditionally convergent.  That is, given any $\alpha\in{\bf R}$ (including
infinity) there is an appropriate reordering of the terms in (8) for which
$d\sigma/d\Omega\ge\alpha$.  Contributions to the differential
cross-section from trajectories not on $\Sigma_{12}$ will not change
this fact.  Since there appears to be no natural ordering for classical
paths, there is no well-defined differential cross-section at
$\theta=\pi/2$.

\smallskip
We have argued that there is no consistent quantisation of the two
$SU(2)$ magnetic monopole dynamical system compatible with the
correspondence principle.  In this context, it is now interesting
to mention a recent numerical study [TRA] of the low-energy classical
dynamics of $SU(2)$ magnetic monopoles and dyons using the geodesic
approximation (equations (2)) to obtain: one, the classical magnetic
monopole differential cross-section, and, two, the two dyon (magnetic
monopole with an electric charge) interaction escape plot.  Surprisingly,
it is found that the classical cross-section at $\theta=\pi/2$ in the
centre-of-mass frame is in excellent agreement with the approximate
quantum mechanical behaviour predicted for the s-wave in a partial wave
analysis of the Gibbons-Manton Schr\"odinger equation [TRA,S].  Moreover,
there appears to be a natural way to define isolated dyonium bound states
from the classical `escape plots', by requiring that bound states be
bounded in both time directions.  These results raise the following
questions concerning the foundations of classical and quantum soliton
dynamics in gauge theories.  Can the classical low-energy soliton dynamics
encoded within Yang-Mills-Higgs theory account for some of the expected
quantum mechanical behaviour?  What is the dyonium energy spectrum?  What is
the statistical distribution of the dyonium energy levels?  These are
questions for further research.

\vskip 0.5in
\noindent
\centerline {\bf References}
\medskip
\item{[AH]} M.F. Atiyah, N.S. Hitchin, The Geometry and Dynamics of
Magnetic Monopoles (Princeton U.P., Princeton, 1989).
\smallskip
\item{[M]} N.S. Manton, {\it Phys. Lett.} B {\bf 110} (1982) 54.
\smallskip
\item{[GM]} G. Gibbons, N. Manton, {\it Nucl. Phys. B} {\bf 274}
(1986) 183.
\smallskip
\item{[P]} P. Pechukas, {\it Phys. Rev.} {\bf 181} (1969) 166.
\smallskip
\item{[S]} B.J. Schroers, {\it Nucl. Phys. B} {\bf 367} (1991) 177.
\smallskip
\item{[TR1]} M. Temple-Raston, {\it Phys. Lett.} B {\bf 206}
(1988) 503.
\smallskip
\item{[TR2]} M. Temple-Raston, {\it Phys. Lett.} B {\bf 213}
(1988) 168.
\smallskip
\item{[TR3]} M. Temple-Raston, {\it Nucl. Phys.} B {\bf 313}
(1989) 447.
\smallskip
\item{[TRA]} M. Temple-Raston, D. Alexander (1992) Concordia preprint
 92-3, hep-th/9209063.

\bye